

\magnification1200
\font\tenrm=cmr10 
\font\figfont=cmr8

\baselineskip13pt plus 2pt minus 3pt
\input psfig
\newcount\countfig
\input epsf
\def\putfig#1#2#3{
\goodbreak\midinsert
\global\advance\countfig by 1
\xdef#1{\number\countfig}
{\figfont \noindent FIG.\ \number\countfig. #3}
\endinsert}
\newcount\counteqn
\def\eqnum#1{\global\advance\counteqn by 1
\eqno(\number\countsec.\number\counteqn)
\xdef#1{(\number\countsec.\number\counteqn)}}
\def\eqalnum#1{\global\advance\counteqn by 1
&(\number\countsec.\number\counteqn)
\xdef#1{(\number\countsec.\number\counteqn)}}

\def\belabel#1{$$\global\advance\counteqn by 1
\xdef#1{(\number\countsec.\number\counteqn)}
}
\def\ee{\eqno(\number\countsec.\number\counteqn)$$}
\def\be{$$\global\advance\counteqn by 1}

\def\bea#1#2{\global\advance\counteqn by 1
\xdef#1{(\number\countsec.\number\counteqn)}
$$\eqalignno{#2 &(\number\countsec.\number\counteqn)\cr}$$}

\def\beanolabel#1{\global\advance\counteqn by 1
$$\eqalignno{#1 &(\number\countsec.\number\counteqn)\cr}$$}
\newwrite\refout
\immediate\openout\refout=reference.txa
\def\ref#1#2#3{
	\xdef#1{#3}
	\immediate\write\refout{\par\hangindent=.725 truein \hangafter=1\ #2}
		}

\def\rc#1{\ (#1)}
\newcount\countsec
\def\newsec#1{\global\advance\countsec by 1
\goodbreak\medskip\goodbreak{\parindent0pt{\bf \number\countsec. #1}}
\nobreak
\medskip
\nobreak\counteqn = 0
}
\def\vs{\noalign{\vskip5pt}}
\def\me{m_e}
\def\te{T_e}
\def\ti{\tau_{\rm initial}}
\def\tci#1{n_e(#1) \sigma_T a(#1)}
\def\tr{\tau_R}
\def\dtr{\delta\tau_R}
\def\dd{\tilde\Delta^{\rm Doppler}}

\def\clsw{C_l^{\rm Sachs-Wolfe}}
\def\cldop{C_l^{\rm Doppler}}
\def\Dt{\tilde{\Delta}}
\def\mut{\mu}
\def\vt{\tilde v}
\def\hp{ {\bf \hat p}}
\def\sdv{S_{\delta v}}
\def\svv{S_{vv}}
\def\bvt{\tilde{\bv}}
\def\delt{\tilde{\delta_e}}
\def\cos{{\rm cos}}
\def\nn{\nonumber \\}
\def\bq{ {\bf q} }
\def\ba{ {\bf p} }
\def\bap{ {\bf p'} }
\def\bqp{ {\bf q'} }
\def\bp{ {\bf p} }
\def\bpp{ {\bf p'} }
\def\bk{ {\bf k} }

\def\bv{ {\bf v} }

\def\bkp{ {\bf k'} }
\def\gq{ g(\bq)}
\def\gqp{ g(\bqp)}
\def\fp{ f(\bp)}
\def\h#1{ {\bf \hat #1}}
\def\fpp{ f(\bpp)}
\def\fz{f^{(0)}(p)}
\def\fpz{f^{(0)}(p')}
\def\f#1{f^{(#1)}(\bp)}
\def\fps#1{f^{(#1)}(\bpp)}
\def\dq{ {d^3\bq \over (2\pi)^32E(\bq)} }
\def\dqp{ {d^3\bqp \over (2\pi)^32E(\bqp)} }
\def\dpp{ {d^3\bpp \over (2\pi)^32E(\bpp)} }
\def\dtq{ {d^3\bq \over (2\pi)^3} }

\def\part#1;#2 {\partial#1 \over \partial#2}
\def\deriv#1;#2 {d#1 \over d#2}

\def\Dtwo{\tilde\Delta^{(2)}}

\def\delpp{\delta(p-p')}

\def\tc{\tau_0}
\def\DD{\langle|\Delta(k,\mu,\tau_0)|^2\rangle}
\def\DDL{\langle|\Delta(k=l/\tc,\mu)|^2\rangle}
\def\bkpp{{\bf k''}}
\def\kmkp{|\bk-\bkp|}
\def\kmkpsq{k^2+k'^2-2kk'x}

\font\titlefont=cmbx10 at 14.4truept
\font\namefont=cmr12

\ref\JEREMY{Bernstein, J. 1988, {\it Relativistic Kinetic Theory}
(Cambridge University).}{Bernstein 1988}
\ref\myself{Bernstein, J. $\&$ Dodelson, S. 1990, {\it Phys. Rev. D}, {\bf 41},
354.}{Bernstein $\&$ Dodelson 1990}
\ref\BELM{Bond, J. R., Efstathiou, G., Lubin, P. M., $\&$ Meinhold, P. R. 1991,
{\it Phys. Rev. Lett.}, \ \ \ {\bf 66}, 2179.}{Bond {\it et al.} 1991}
\ref\BEFS{Bond, J. R. $\&$ Efstathiou, G. 1987, {\it MNRAS}, {\bf 226},
655.}{Bond $\&$ Efstathiou 1987}
\ref\CRITTENDEN{Crittenden, R., Bond, J. R., Davis, R. L., Efstathiou, G., $\&$
Steinhardt, P. J. 1993,\ \ \ {\it Phys. Rev. Lett.}, {\bf 71}, 324.}{Crittenden
{\it et al.} 1993}
\ref\TENERIFE{Davies, R. D., Lasenby, A. N., Watson, R. A., Daintree, E. J.,
Hopkins, J., Beckman, J., Sanchez-Almeida, J., $\&$ Rebolo, R. 1987 {\it
Nature}, {\bf 326}, 462.}{Davies {\it et al.} 1987}
\ref\DDM{Dodelson, S. $\&$ Jubas, J. M. 1992, {\it Phys. Rev. D}, {\bf 45},
1076.}{Dodelson $\&$ Jubas 1992}
\ref\US{Dodelson, S. $\&$ Jubas, J. M. 1993, {\it Phys. Rev. Lett.}, {\bf 70},
2224.}{Dodelson $\&$ Jubas 1993}
\ref\DOROSH{Doroshkevich, A. G., Zel'dovich, Ya. B., $\&$ Sunyaev, R. A. 1978,
{\it Sov. Astr.}, {\bf 22}, 523.}{Doroshkevich, Zel'dovich, $\&$ Sunyaev 1978}
\ref\Efstathiou{Efstathiou,G. 1988, in {\it Large Scale Motions in the
Universe.  A Vatican
 Study Week}, \qquad \quad \ eds V. C. Rubin $\&$ G. V. Coyle
(Princeton:Princeton University
 Press), p299.}{Efstathiou 1988}
\ref\EBW{Efstathiou, G., Bond, J. R., $\&$ White, S. D. M. 1992, {\it MNRAS},
{\bf 258}, 1P.}{Efstathiou {\it et al.} 1992}
\ref\GAIER{Gaier, T., Schuster, J., Gundersen, J., Koch, T., Seiffert, M.,
Meinhold, P., $\&$ Lubin, P. 1992 {\it Ap. J.}, {\bf 398}, L1.}{Gaier {et al.}
1992}
\ref\SCOTT{Hu, W., Scott, D., $\&$ Silk, J. 1993, preprint.}{Hu, Scott, $\&$
Silk 1993}
\ref\KAISER{Kaiser, N. 1984 {\it Ap. J.}, {\bf 282}, 374.}{Kaiser 1984}
\ref\KOMPANEETS{Kompaneets, A. S. 1957, {\it J. E. T. P.}, {\bf 4},
730.}{Kompaneets 1957}
\ref\MARTINEZ{Martinez-Gonzalez, E., Sanz, J.L., $\&$
Silk, J. 1992, {\it Phys. Rev. D}, {\bf 46}, 4193.}{Martinez-Gonzalez, Sanz,
$\&$ Silk 1992}
\ref\MAX{P. R. Meinhold, {\it et al.} 1993, {\it Ap. J.}, 409, L1.}{Meinhold
{\it et al.} 1993}
\ref\PEEBLES{Peebles, P. J. E. 1987 {\it Ap. J.}, {\bf 315}, L51.}{Peebles
1987}
\ref\PEEBYU{Peebles, P. J. E. $\&$ Yu, J. T. 1970, {\it Ap. J.}, {\bf 162},
815.}{Peebles $\&$ Yu 1970}
\ref\Readhead{Readhead, A. C. S. {\it et al.}, 1989, {\it Ap. J.}, {\bf 346},
566.}{Readhead {et al.} 1989}
\ref\SACHS{Sachs, R. K. $\&$ Wolfe, A. M. 1967, {\it Ap. J.}, {\bf 147},
73.}{Sachs $\&$ Wolfe 1967}
\ref\COBE{Smoot, G. F. {\it et al.} 1992, {\it Ap. J.}, {\bf 396},
L1.}{Smoot {et al.} 1992}
\ref\VISHNIAC{Vishniac, E. 1987, {\it Ap. J.}, {\bf 322}, 597.}{Vishniac 1987}
\ref\WILSON{Wilson, M. L. $\&$ Silk, J. 1981, {\it  Ap. J.}, {\bf 243},
14.}{Wilson $\&$ Silk 1981}
{
\nopagenumbers
\rightline{\hfill FERMILAB-Pub-93/242-A}
\rightline{\hfill August 1993}
\vskip 0.5truein
\titlefont
\centerline{RE-IONIZATION AND ITS IMPRINT ON THE}
\centerline{COSMIC MICROWAVE BACKGROUND}
\medskip\medskip\medskip
\noindent
\namefont
\centerline{Scott Dodelson$^1$ and Jay M. Jubas$^{2}$}
\tenrm
\medskip
\centerline{$^1$NASA/Fermilab Astrophysics Center}
\centerline{Fermi National Accelerator Laboratory}
\centerline{P.O. Box 500, Batavia, IL 60510}

\medskip
\centerline{$^2$Department of Physics}
\centerline{Massachusetts Institute of Technology}
\centerline{Cambridge, MA 02139}

\vskip 0.5truein \baselineskip=20truept plus 1pt minus 2pt

\centerline{ABSTRACT}
Early reionization changes the pattern of anisotropies expected in the cosmic
microwave background. To explore these changes, we derive from first principles
the equations governing anisotropies, focusing on the interactions of photons
with electrons.
Vishniac (1987) claimed that second order terms can be large in a re-ionized
Universe, so we derive equations correct to second order in the perturbations.
There are many more second order terms than were considered by Vishniac. To
understand the basic physics involved, we present a simple analytic
approximation to the first order equation.  Then turning to the second order
equation, we show that the Vishniac term is indeed the only important one.
We also present numerical results for a variety of ionization histories [in a
standard cold dark matter Universe] and show quantitatively how the signal in
several experiments depends on the ionization history. The most pronounced
indication of a re-ionized Universe would be seen in very small scale
experiments; the expected signal in the Owens Valley experiment is smaller by a
factor of order ten if the last scattering surface is at a redshift $z\simeq
100$ as it would be if the Universe were re-ionized very early. On slightly
larger scales, the expected signal in a re-ionized Universe {\it is} smaller
than it would be with standard recombination, but only by a factor of two or
so. The signal is even smaller in these experiments in the intermediate case
where {\it some} photons last scattered at the standard recombination epoch.
\footnote{}{
Submitted to {\it Astrophysical Journal}}
\vfill\eject}
\pageno=1
\newsec{Introduction}
Measurements of the cosmic microwave background (CMB) are very powerful probes
of  theories of structure formation. Even the so-called ``medium-angle'' CMB
experiments are looking at scales larger than those in the largest redshift
surveys. This means that CMB observations are some of the purest observations
in the field: more than any other type of experiment they sample the primordial
spectrum of fluctuations. We can therefore hope to use the CMB experiments to
answer many questions about primordial fluctuations: Are they Gaussian? What is
the spectral index? Are they adiabatic or isocurvature? Do gravity waves play
any role?

These are profound questions and the possibility that we might answer them is
one of the great promises of observational cosmology. However,
there are several flies in the ointment. Perhaps most disturbing is foreground:
our galaxy has lots of dust, synchrotron and free-free emission. All of these
can get in the way of detecting the purely cosmic radiation. Another potential
barrier to inferring the primordial spectrum from CMB measurements -- and the
one we wish to focus on here -- is the possibility of {\it reionization}. In
the standard cosmology the electrons and protons ``recombine'' at redshift
$z\simeq 1100$, thereby cutting off contact between photons and matter.
However, it is possible that hydrogen was ``re-ionized'' at a later epoch. The
subsequent contact between photons and free electrons changes the pattern of
anisotropies in the CMB.  Thus {\it we can use the CMB to infer the ionization
history of our
Universe}. As we have said, knowing this history is crucial for the sake of
determining the primordial spectrum, but it is also
interesting in of itself. For example, if we determine that the Universe was
ionized at a redshift $z\sim 100$, we will have learned something very
useful about structure at that epoch.

In this paper, we would like to quantify these statements with particular
attention to experiments. The simplest way to discuss anisotropies in the CMB
is in $l-$ space\rc\BELM. That is,
we expand $\delta T(\theta,\phi)/T = \sum_{lm} a_{lm} Y_{lm}(\theta,\phi)$ and
define $C_l \equiv \langle \vert a_{lm}\vert^2 \rangle$, where the angular
brackets denote ensemble averages. Each theory has its own set of predicted
$C_l$'s and therefore predicts that a given experiment will observe a variance
$$ \langle (\Delta T/T)^2_{\rm expt} \rangle
	= \sum_{l=2}^\infty {2l+1\over 4\pi} C_l W_{l, {\rm expt}}
 \eqnum\cldef $$
where $W_l$ is the window function appropriate to that experiment.
\putfig\FILTERS{filters.eps}{Window functions for several experiments currently
searching for anisotropies in the cosmic microwave background. The curve
labeled SP91 denotes the filter for the ACME telescope for the nine-point scan
of 1991
\rc\GAIER\ in the South Pole; TENERIFE the one used by Davies {et al.} (1987)
at Tenerife; MAX the filter used by Meinhold {et al.} (1993) in the region of
$\mu$-Pegasus; OVRO the filter used at Owens Valley \rc\Readhead. For reference
the COBE \rc\COBE\ filter is also plotted.}
Figure \FILTERS\
shows the window functions of a variety of experiments. For large $l$, the
expected variance in a given experiment is seen from Eq. \cldef\ to be roughly
$1/2\pi$ $\int d\ln(l)\ (l^2C_l) W_l $. So the quantity $(l^2C_l)$ is
convolved
with the window function to give the expected variance. Figure 2 plots
the $l^2C_l$ predicted by cold dark matter (CDM) with standard
ionization history. The peak at
$l=200$ is probed by several of the experiments shown in Figure \FILTERS.

\putfig\CLCDM{clcdm.eps}{The moments predicted by CDM with standard
recombination. The quantity plotted $l(l+1)C_l/6C_2$ is equal to one at low $l$
where only the Sachs-Wolfe effect is significant. In the post-COBE
era, we know that $l(l+1)C_l/6C_2 = l(l+1)C_l/(2.4\times 10^{-5})^2$.}

We want to know how the $C_l$'s change if the Universe is re-ionized. Is there
still a peak at $l=200$? Does the amplitude change? Does the peak move, perhaps
to lower or higher $l$? Or both? To answer these questions we derive the
fundamental Boltzmann equation governing the interaction between photons and
electrons, the interaction that, as we will see, is responsible for the peak in
Fig. 2. This derivation is presented in sections
in what we think is a very systematic
way. One of the advantages of this systematic
treatment is it will enable us to
pick up not only the linear terms [which of course have already been derived
many times\rc{\PEEBYU;\ \WILSON}] but also the second order terms. These are of
particular interest
because Vishniac (1987) has analyzed one such second order term and
concluded that in
a reionized Universe, anisotropies are {\it generated} at small scales [high
$l$]. By introducing this systematic treatment of the Boltzmann equation, we
will be able to see if there are other second order terms that are as large as
those considered by Vishniac, and if cancellations occur.

First though, in section 4, we solve the first order equations. The focus is on
two possible ionization histories: the standard scenario wherein electrons and
protons recombine at $z\sim 1100$ with no further ionization and the opposite
extreme wherein electrons remain free throughout the whole history of the
Universe. In addition to presenting
the results of a full numerical treatment,
we also
solve the equation [or an approximation to it] analytically to gain insight
into the location of the peak in $l-$ space. We find that this peak shifts to
{\it lower} $l$ if the Universe is re-ionized early enough. If the Universe
never recombines, then there is a peak at $l\sim 50$. The signal in an
experiment with a filter centered around this value of $l$ would see a {\it
larger} signal in a completely re-ionized Universe than in one with standard
recombination. We also show why $l^2C_l$ falls off at high $l$. This will lead
into a
discussion of the second order terms expected to be important.
Sections 5 and 6 use this information to complete the derivation of the second
order Boltzmann equation and present the solution. Our techniques are similar
to those of Efstathiou (1988).
In particular we will recapture
Vishniac's result and show that the term he identified is indeed the dominant
one. Thus {\it we support Vishniac's conclusions about second order effects.}
Finally in section 7, we get more specific about experiments, presenting
the predicted $ \langle (\Delta T/T)^2_{\rm expt} \rangle$ in a variety of
experiments for a variety of ionization histories.

One final introductory comment: The signal from the cosmic background depends
on many parameters even in a model as simple as CDM. For example, the spectral
index $n$ need not be equal to one, and even a slight deviation has dramatic
implications for the small and medium scale measurements. The signal in a given
experiment also depends\rc{\BELM; \US} on $\Omega_B$, the fraction of critical
density in baryons today, and $h$ which parametrizes the Hubble constant today
$(H_0 = 100h
{\rm km} {\rm sec}^{-1} {\rm Mpc}^{-1})$. And of course, models other than CDM
give different predictions and have different sets of parameters to fiddle
with. Here we are focusing only on the impact of different ionization
histories.
It therefore makes sense to fix all these other parameters and play with the
one variable of interest: the ionization history. Accordingly, we will focus
only on a cold dark matter dominated Universe with Harrison-Zel'dovich spectrum
$(n=1)$ and set $h=1/2$ and $\Omega_B = 0.05$. Several other groups\rc{\BEFS;
\Efstathiou; \SCOTT}\ have recently considered the effects of re-ionization for
other cosmologies, in particular the minimal isocurvature model proposed by
Peebles (1987).

\newsec{Compton Collision Term: A General Derivation}

The photon spectrum is governed
by the Boltzmann equation:
\belabel\fund
 {d\over dt} f({\bf x}, \ba, t) = C({\bf x}, \ba, t) .
\ee
Here $f$ is the photon occupation number, a function of momentum $\ba$,
position ${\bf x}$, and time $t$. The collision term, $C$, also depends
on these variables. Theoretically, it includes contributions from all
scattering processes, although in practice only Compton scattering
off free electrons need be considered.

In the absence of collisions [$C=0$], Eq.
\fund\ says simply that
photons travel freely along geodesics. For example, in a Robertson-
Walker background, the left hand side of Eq. \fund\ becomes
\belabel\berw
{df\over dt} = \left\{ {\partial\over \partial t} + {\ba\over p}\cdot
{\partial\over \partial {\bf x}} - H p {\partial
						\over\partial p}
\right\} f({\bf x},\ba,t)
\ee
where $H$ is the Hubble rate and $p=\vert\ba\vert$. If we are interested in
large scale
anisotropies, we must do better: we must account for the
perturbed metric when expanding $d/dt$ in terms of partial derivatives.
Such an account leads to the Sachs-Wolfe effect\rc\SACHS. [Recently,
Martinez-Gonzalez, {\it etal.} (1992) have taken this program a step
further and considered second order effects in the perturbations to the
metric.]

In this paper we will focus on the right hand side of Eq. \fund: the
collision term. This term governs small scale anisotropies and spectral
distortions. In the limit of completely elastic collisions, the right hand side
vanishes. Typically, in the
regime of interest, very little energy is imparted from electrons to
photons in a collision, so this limit
is a good approximation. To get non-zero effects, therefore, we simply
need to expand the right hand side systematically in powers of the energy
transfer.

The starting point then is the collision term corresponding to the process
$$e(\bq) \gamma(\ba) \leftrightarrow e(\bqp) \gamma(\bap).$$
To calculate this, we find the matrix element $M$ for the process, square it,
weight it by the occupation numbers of the particles and integrate over
all other momenta, $\bq,\bqp,\bap$.
Therefore,
\bea\ecoll{
C(p) &= {1 \over p} \int \dq\dqp\dpp(2\pi)^4
\delta^4(q+p-q'-p')|M|^2 \cr
\vs
&\qquad\times \Bigg[ \gqp\fpp\bigg(1+\fp\bigg)\quad
	-\quad \gq\fp\bigg(1+\fpp\bigg) \Bigg]}
where $E(q) = \sqrt{q^2 + m^2}$ and the delta function enforces energy-momentum
conservation.
The last line in Eq. \ecoll\ contains the
distribution functions, $f$ that of the photons and $g$ the electrons.
We have dropped the Pauli suppression factors $1-g$, since
in all realistic cosmological scenarios, $g$ is very small. We don't know $f$
[that is what we are trying to solve for] but we do know $g$: Due to the fast
rate of Coulomb collisions, the electrons are kept in thermal equilibrium, so
\belabel\edist
g(\bq) = n_e \left( {2\pi\over\me\te} \right)^{3/2} \exp\left\{ {-(\bq -
\me\bv)^2\over 2\me\te} \right\},
\ee
where $\bv$ is the velocity of the electrons and the normalization comes from
requiring that $\int d^3q\ g/(2\pi)^3 = n_e$, the electron density.
Before beginning the expansion, we can trivially do the $\bqp$ integration in
Eq.~ \ecoll\ by using the three dimensional delta function. This leaves
\bea\ert{
C(p)&={1 \over 8\pi p} \int dp' p' {d\Omega' \over 4 \pi}
\int \dtq {|M|^2 \over E(\bq)E(\bp+\bq-\bpp)}    \delta\biggl(p+E(\bq) -p'
-E(\bp+\bq-\bpp) \biggr)
\cr\vs
&\qquad\times
\Bigg[ g(\bp+\bq-\bpp)\fpp\bigg(1+\fp\bigg) \quad-\quad
 \gq\fp\bigg(1+\fpp\bigg) \Bigg]}

As mentioned above, the energy transfer, $E(\bq) - E(\bq+\bp-\bpp)$, is small
compared with the typical thermal energies which are of order $T$. In fact, the
energy difference is of order $E(\bq) - E(\bq+\bp-\bpp) \simeq
(\bpp-\bp)\cdot \bq/\me = {\cal O}(Tq/\me)$. Thus our expansion parameter, the
energy difference over the temperature, is actually $q/\me$. The electron
momentum has two sources: the bulk velocity ($q=\me v$) and the thermal motion
($q\sim\sqrt{\me T}$). Thus, an expansion in $q/\me$ is necessarily an
expansion in $v$ and $\sqrt{T/\me}$. At the end of our expansion we will have a
first order source term linear in $v$ and a host of second order terms
quadratic in $v$ and $\sqrt{T/\me}$.

The
strategy now is to expand everything -- energies,
squared matrix element,
delta function
and distribution functions -- using the energy transfer
as an expansion parameter. The Boltzmann distribution expands to
\bea\gexp{
g(\bp+\bq-\bpp) &= \gq \Biggl\{1-{(\bp-\bpp) \cdot (\bq-m\bv)
\over m_eT_e} -
{(\bp-\bpp)^2 \over 2m_eT_e} \cr
&+ {1 \over 2} \biggl[{(\bp-\bpp)
\cdot (\bq-m\bv) \over
m_eT_e}\biggr]^2 + \ldots \Biggr\}
}
The second term in brackets is first order in the perturbative quantities
while the last two are second order.
Meanwhile the delta function expands to
\bea\dexp{
\delta\biggl(p&+E(\bq)-p'-E(\bp+\bq-\bpp)
\biggr) =
\delta(p-p') +
{(\bp-\bpp) \cdot \bq \over m_e} {\part \delpp;p' } \cr
\vs&+
{(\bp-\bpp)^2 \over 2m_e}
{\part \delpp;p' }
+ {1 \over 2} \biggl[{(\bp-\bpp)
\cdot \bq \over m_e} \biggr]^2
{\partial^2 \delpp \over \partial p'^2} + \ldots
}
The derivatives of the Dirac delta functions here will ultimately be handled
by integration by parts. [Bernstein (1988) introduced this approach to derive
the Kompaneets
equation which, as will see in section 5, is a special case of the general
second order equation.]
Finally, we expand the photon distribution function,
\belabel\fexp
f=f^{(0)}(p)+f^{(1)}(\bp)+f^{(2)}(\bp)\ee
where $f^{(0)}(p)$ is the zero order photon distribution [typically Planckian]
which of course depends only on
the magnitude of $p$. We can use these three expansions to rewrite Eq. \ert\
as
\bea\ecolexp{
C(p)&={1 \over 8\pi p} \int dp' p' {d\Omega' \over 4 \pi}
\int \dtq {|M|^2 g(\bq)\over E(\bq)E(\bp+\bq-\bpp)}
\Bigg\{
\delta(p-p') +
{(\bp-\bpp) \cdot \bq \over m_e} {\part \delpp;p' } \cr
\vs&\qquad\qquad+
{(\bp-\bpp)^2 \over 2m_e}
{\part \delpp;p' }
+ {1 \over 2} \biggl[{(\bp-\bpp)
\cdot \bq \over m_e} \biggr]^2
{\partial^2 \delpp \over \partial p'^2} \Bigg\}\cr\vs
&\times
\Bigg\{ \left[ \fpz - \fz \right]
	+ \left[ \fps1 - \f1 - \fpz(1+\fz){ (\bp - \bpp)\cdot (\bq - \me\bv)
	\over \me\te} \right]\cr\vs
&+ \bigg[ \fps2-\f2 - { (\bp - \bpp)\cdot (\bq - \me\bv)
	\over \me\te}\left( \fps1(1+\fz) + \fpz\f1\right) \cr\vs
& + \fpz(1+\fz) \left( {-(\bp - \bpp)^2\over 2\me\te} +
{1\over2}\Big( { (\bp - \bpp)\cdot (\bq - \me\bv)
	\over \me\te} \Big)^2 \right) \bigg] \Bigg\}.
}
This equation looks like a mess, and we have not even expanded all the
quantities on the first line yet. It turns out though that the hard part is
over. We now recognize that the zero order term in Eq. \ecolexp, i.e.
the one we get when multiplying together the first terms in each of the curly
brackets, vanishes. That is, $\delta(p-p')(\fz-\fpz)$ is zero. Therefore
only terms of first order remain after multiplying all the terms in the two
curly brackets. This means that [since we are interested only in terms up to
second order] we only have to keep first order terms when we expand the matrix
element and the energies on the first line. In fact to first order the energies
can be simply replaced by $\me$, and the matrix element squared is
\belabel\mexp
\vert M\vert^2 = 6 \pi \sigma_T m_e^2
\left(
(1+{\rm cos}^2\theta) -
2{\rm cos}\theta(1-{\rm cos}\theta)\bq\cdot( \h p + \h p')/\me
+\ldots \right)
\ee
where ${\rm cos}\theta = \hat \bp \cdot \hat{\bpp}$ and $\sigma_T$ is the
Thomson cross section.
To simplify things further, we can explicitly do the $d^3q$ integral
by using $\langle g \rangle = n_e$; $\langle g \bq\rangle =  n_e \me\bv$;
and $\langle g q_iq_j\rangle = \delta_{ij}n_e \me \te + n_e \me^2v_iv_j,$
where $\langle A \rangle = \int d^3q\ A/(2\pi)^3$.
Then we find
\bea\allcoll{
C(\bp) =& {3n_e\sigma_T\over 4p}\int dp' p' {d\Omega' \over 4 \pi}
\bigg[ c^{(1)}(\bp,\bpp) + c^{(2)}_K(\bp,\bpp) + c^{(2)}_\Delta(\bp,\bpp)
\cr\vs
&\qquad\qquad\qquad\qquad\qquad + c^{(2)}_{\Delta v}(\bp,\bpp) +
c^{(2)}_{vv}(\bp,\bpp)   \bigg]
}
where the first order integrand is
\bea\cone{
c^{(1)}(\bp,\bpp) &= (1+\cos^2\theta)\Bigg(
\delta(p-p') (\fps1-\f1)
\cr\vs &\qquad\qquad\qquad\qquad
+\quad (\fpz-\fz) (\bp-\bpp)\cdot\bv {\partial \delta(p-p')
			\over \partial p'} \Bigg);
}
and we have separated the second order terms into four parts. The first set of
these contribute to what is referred to as the Kompaneets
equation\rc\KOMPANEETS\ describing spectral distortions to the CMB\rc\myself:
\bea\ckom{
c^{(2)}_K(\bp,\bpp) &= \left(1+\cos^2\theta\right) {(\bp-\bpp)^2\over2\me}
	\Bigg[  \left( \fpz-\fz\right) \te {\partial^2 \delta(p-p')
			\over \partial p'^2}
\cr\vs &\qquad\qquad\qquad
		- \left(\fpz+\fz+2\fpz\fz\right) {\partial \delta(p-p')
			\over \partial p'} \Bigg]
\cr\vs
 + &{2(p-p')\cos\theta(1-\cos^2\theta)\over\me}
	\Bigg[ \delta(p-p') \fpz(1+\fz)
\cr\vs &\qquad\qquad\qquad\qquad\qquad -
		\left(\fpz-\fz\right){\partial \delta(p-p')
			\over \partial p'} \Bigg]
.}
There is also the simple damping term
\belabel\edamptwo
c^{(2)}_\Delta(\bp,\bpp) =\left(1+\cos^2\theta\right)
		 \delta(p-p')(\fps2 - \f2) ;
\ee
a set of terms coupling the photon perturbation to the velocity
\bea\edelv{
 c^{(2)}_{\Delta v}(\bp,\bpp)=&
	\left(1+\cos^2\theta\right) \left(\fps1 - \f1\right)
\Bigg[  \left(1+\cos^2\theta\right)(\bp-\bpp)\cdot\bv {\partial \delta(p-p')
			\over \partial p'}
\cr\vs &\qquad\qquad\qquad\qquad -
\ 2\cos\theta(1-\cos\theta)
	\delta(p-p')
		(\h p + \h p')\cdot \bv
\Bigg];
}
and finally a set of source terms quadratic in the velocity:
\bea\ctwovv{
c^{(2)}_{vv}(\bp,\bpp) =&
	\left(\fpz-\fz\right)\ (\bp-\bpp)\cdot\bv
\Bigg[
	\left(1+\cos^2\theta\right) {(\bp-\bpp)\cdot\bv\over2}
		 {\partial^2 \delta(p-p')
			\over \partial p'^2}
\cr\vs &\qquad\qquad\qquad\qquad -\
2\cos\theta(1-\cos\theta)
(\h p + \h p')\cdot \bv{\partial \delta(p-p')
			\over \partial p'}
.\Bigg]
}
We'll see in the next section that the first order terms, those in Eq. \cone,
reduce to the standard first order equation
once the $p'$ and $\Omega'$ integrals are done. There are a lot of second order
terms. As we have mentioned, one subset of these terms has already been
studied: those leading to the Kompaneets equation\rc\KOMPANEETS. We know that
Vishniac analyzed a second order term. Which one is it here? It turns out that
he looked at none of the second order terms in $c^{(2)}$ [or at least he didn't
write a paper about any of them]. Rather, he got yet another second order term
by expanding the electron density as
$n_e = \bar{n_e}(1+\delta_e)$ and then multiplying $\delta_e$ by the first
order terms in Eq. \cone.

\newsec{First Order Equation}

In this section, we derive the well-known first order equation
coupling photons and electrons. To do this, we need focus only on the terms in
Eq.
\cone. For the angular integrals, we choose the polar axis to lie
along the direction of the electron velocity, so that azimuthal symmetry
is maintained. Then $\mu'$ is the polar
angle defined by $\mu' \equiv \hat \bv \cdot \hat \bpp$; we also define
$\mu \equiv \hat \bv \cdot \hat \bp$.
Thus we have
\def\pdp{\h p\cdot\h p'}
\bea\econein{
C^{(1)}&(\bp)= {3n_e \sigma_T\over 4p}\ \int dp' p' {d\Omega' \over 4
\pi} \ c^{(1)}(\bp,\bpp) \cr\vs
&={3 n_e\sigma_T \over 4 p} \int_0^\infty dp' p'
\Bigg[
	\delta(p-p') \int_{-1}^1 {d\mu'\over2} \ (\fps1-\f1)
			\int_0^{2\pi} {d\phi'\over2\pi}(1+(\pdp)^2)\cr\vs
&	+ v(\fpz-\fz){\partial \delta(p-p')
			\over \partial p'}\int_{-1}^1 {d\mu'\over2} \
		(p\mu-p'\mu')\int_0^{2\pi} {d\phi'\over2\pi}(1+(\pdp)^2)
\Bigg]
,}
where $v\equiv\vert \bv\vert$. The dot product $\pdp$ is a function of both
$\mu'$ and $\phi'$. But
$\fps1$ depends only on $p'$ and $\mu'$, not on $\phi'$ because of the
azimuthal symmetry. [This is perhaps too strong a statement. For many metrics
considered by cosmologists, there {\it is} an azimuthal symmetry so $f^{(1)}$
depends only on the polar angle. There are cosmologies though wherein this
symmetry is not maintained, so $f^{(1)}$ could well depend on the azimuthal
angle. This happens for example when tensor perturbations \rc\CRITTENDEN\ or
cosmic strings are
present. We will restrict our analysis to cases where the symmetry exists.]
To do the $\phi'$ integral we first rewrite the integrand in terms of Legendre
polynomials
\be
 1 + (\pdp)^2 = { 4 \over 3} \left( 1 + {1\over2}P_2
(\pdp) \right)
\ee
By the
addition theorem of spherical harmonics,
\be
P_2(\hat \bp \cdot \hat \bpp)  =
\sum_{m=-2}^{m=2}
{(2-m)! \over (2+m)!} P_2^m(\hat \bp \cdot \hat \bv)
P_2^m(\hat \bpp \cdot \hat \bv)e^{im(\phi'-\phi)}
\ee
Thus,
\be
\int {d \phi' \over 2 \pi} P_2(\hat \bp \cdot \hat \bpp)
= P_2(\hat \bp \cdot \hat \bv)P_2(\hat \bpp \cdot \hat \bv) \equiv
P_2(\mu)P_2(\mu')
\ee
To do the $\mu'$ integrals, it is useful to define the moments of the
distribution function:
\belabel\defmom
f_l(p)=\int_{-1}^1 {d\mu \over 2} P_l(\mu) f(p,\mu).
\ee
Now we can use the orthonormality of the Legendre polynomials to write
\belabel\ortho
\int_{-1}^1 {d\mu' \over 2} \big(1+P_2(\mu)P_2(\mu')\big)
\left[f^{(1)}(\bpp) - f^{(1)}(\bp)\right] =
f^{(1)}_0(p') + {1 \over 2}f^{(1)}_2(p')P_2(\mu)-f^{(1)}(p,\mu)
\ee
The notation at this stage is a bit confusing, so let's restate it:
the superscript refers to the order of perturbation theory. Here we are
considering the first order correction. The subscript refers to the moment of
the distribution function; note that the last term in Eq. \ortho\ has not been
integrated over,
{\it i.e.} still depends on $\mu$, so it has no subscript.
Eq. \econein\ now becomes
\bea\conemed{
C^{(1)}(\bp)=&{n_e\sigma_T \over p} \int_0^\infty dp' p'
\Bigg[
	\delta(p-p') \left( f^{(1)}_0(p') + {1 \over
		 	2}f^{(1)}_2(p')P_2(\mu)-f^{(1)}(p,\mu) \right)
\cr\vs
&\qquad\qquad\qquad	+ vp\mu\left(\fpz-\fz\right){\partial \delta(p-p')
			\over \partial p'}
\Bigg]
.}
The remaining $p'$ integral can be done, in the first case trivially and in the
second by integrating by parts. Thus our final expression for the collision
term is
\belabel\efirst
C^{(1)}(p,\mu)=n_e\sigma_T \biggl[f^{(1)}_0
+ {1 \over 2}f^{(1)}_2 P_2(\mu)
-f^{(1)} - p { \partial f^{(0)} \over \partial p } \mu v
\biggr].
\ee
It appears as if this collision term is momentum dependent. To get rid of this
illusion, it is useful to define
\belabel\defdel
\Delta(\mu) \equiv - \left[
	{p\over4}\  { \partial f^{(0)} \over \partial p } \right]^{-1}
	f^{(1)}(p,\mu)
.\ee
Then combining the left hand side of the Boltzmann equation from Eq. \berw\
and the right hand from Eq. \efirst, the full first order equation
is
\belabel\efirstorder
\left\{ {\partial\over \partial t} + {\ba\over p}\cdot
{\partial\over \partial {\bf x}} \right\} \Delta
=
n_e\sigma_T
\biggl(\Delta_0 +
 {1 \over 2}
\Delta_2 P_2(\mu) + 4\mu v - \Delta \biggr)
\ee
[The expansion term in Eq. \berw\ simply forces the zero
order distribution function to depend on the {\it comoving} momentum $pa$ where
$a$ is the scale factor. One can then show that the expansion term drops out of
the first order
equation for $\Delta$.]
The subscripts here again refer to the moments of $\Delta(\mu)$
defined just as in Eq. \defmom. Since to first order $\Delta$ is independent of
photon energy, $p$,
$\Delta$ as defined by Eq. \defdel\ is equal to the brightness
defined as
\belabel\altdef
\Delta = {\int dp p^3 f^{(1)} \over \int dp p^3 f^{(0)} }.
\ee

\newsec{First Order Solutions}

Now that we have the equations describing the interactions of photons with
electrons, we can solve them to determine the predicted anisotropies in the
cosmic microwave background. Figure 2 shows the
results of numerically integrating the full set of linear equations starting
with these initial conditions, assuming standard recombination at $z\simeq
1100$. We'd like to do two things in this section:
First, we would like to understand the peak at $l\simeq 200$. If we understand
why it occurs at $l\sim 200$ when the Universe follows a standard ionization
history, then we will be able to understand how this peak shifts when we
consider different ionization histories. Second, we would like to understand
the damping that occurs at $l\sim 1000$. We will see that this damping is due
to the ``finite thickness of the last scattering surface.'' For a re-ionized
Universe, this thickness is much larger, and therefore, damping is apparent
even on very large scales [much smaller $l$]. It turns out that understanding
the physical reason for this damping will give us a clue as to which second
order terms are likely to be significant. [Due to the crudeness of our
approximations, we will not be able to account for the oscillations
that occur as $C_l$ is damped; these are due to acoustic waves at
the time of recombination.]

To understand these features, we will solve a simplified version of Eq.
\efirstorder. First note that $\Delta$ depends on position; to account for
this, we can Fourier transform it:
\belabel\ftrans
\Delta({\bf x},\bp,\tau) = V \int d^3k\ \Dt(k,\mut,\tau) e^{i{\bf
k}_{\rm physical} \cdot {\bf x} },
\ee
where $k_{\rm physical} = k/a$; $k$ being
the comoving wavenumber; $a$ the cosmic scale factor;  $V$ the volume, a factor
which drops out of all physical results. Note that because of
the azimuthal symmetry, $\Dt$ depends only on the magnitude of $k$ and the dot
product $\h k\cdot \h\bp = {\bf {\tilde v}}\cdot \h p =\mut$. Note that the
first equality holds here
since the velocity is irrotational, meaning that the Fourier transform of the
velocity is parallel to ${\bf k}$. We also define the
conformal time
\be
\tau \equiv \int_0^t {dt'\over a(t')}.
\ee
The conformal time today is
\be
\tc = {2\over H_0} = 1.2\times 10^{18}\ {\rm sec}
\ee
since we're assuming a flat, matter, dominated Universe with $h=0.5$.
The first order equation is now
$$ \dot{\Dt} + ik\mut \Dt
		= n_e \sigma_T a \left( 4\mu \vt - \Dt
\right) \eqnum\simpeq $$
where the dot denotes derivative with respect to $\tau$.
The $\Delta_0$
and $\Delta_2$ terms on the right hand side of Eq. \efirstorder\ have been
neglected. That's because we are only interested in the kick that photons
get from the moving electrons [the $4\mu \vt$ term], the so-called {\it
Doppler effect}. The $\Delta_0,\Delta_2$ terms give rise to what's sometimes
called the {\it intrinsic} anisotropy. Also we still have not written down the
terms that represent the perturbation to the metric. These lead to the
Sachs-Wolfe effect. So all we should get out of the simplified Eq. \simpeq\ is
the Doppler effect.

Eq. \simpeq\ is a first order differential equation, whose solution is
$$ \eqalignno{
\Dt(k,\mut,\tau) =& \Dt(k,\mut,\ti)
\exp\left\{ ik\mut (\ti-\tau)
	-\int_{\ti}^\tau d\tau'\tci{\tau'}\right\} \cr
&\qquad + 4\mu \int_{\ti}^\tau d\tau' \tci{\tau'} \vt(k,\tau')
e^{ik\mut(\tau'-\tau)}
	\cr
&\qquad\qquad\qquad\times\exp\left\{
		-\int_{\tau'}^\tau d\tau'' \tci{\tau''}\right\}. \eqalnum\longsol} $$
The first term on the right represents the anisotropies that were initially
present and have persisted until some late time $\tau$. These are damped out by
scattering with electrons, the $\int n_e\sigma_T a d\tau'$ term in the
exponential. If we choose an early enough $\ti$, then this integral is always
large, and so the initial anisotropies are not important today. We can write
Eq. \longsol\ in a more compact form by setting $\ti$ to zero and by defining
the {\it visibility function}
$$ g(\tau,\tau') \equiv \tci{\tau'} \exp\left\{
		-\int_{\tau'}^\tau d\tau'' \tci{\tau''}\right\}. \eqnum\visibl $$
Then we have simply
$$ \Dt(k,\mut,\tau) = 4\mu \int_0^\tau d\tau' \vt(k,\tau') g(\tau,\tau')
e^{ik\mut(\tau'-\tau)} .\eqnum\simpsol $$

The visibility function defined in Eq. \visibl\ has an interesting physical
interpretation. To see this, first note that $\int_0^\infty d\tau'
g(\infty,\tau')
=1$, so $g$ is normalized like a probability density. In fact, that's what it
is: the
probability per unit conformal time that a photon at time $\tau$ was last
scattered at time $\tau'$. The visibility function depends on the free electron
density $n_e$; thus it is particularly sensitive to the ionization history.
Figure 3 shows the visibility function for two ionization histories: (i)
standard recombination at $z\sim 1100$ and (ii) the case where the electrons
remain ionized throughout the whole history of the Universe. In the first case
recombination happens rapidly, so the ``surface of last scattering'' is
centered at $z=1100$ but with a very small width. In the second case, the
surface of last scattering is centered at $z = 100$ but is quite wide. To
simplify Eq. \simpsol\ further, we can approximate $g$ as a Gaussian:
$$ g(\tau_0,\tau') \simeq {1\over  \sqrt{\pi}\dtr}
	\exp\left\{
		- {(\tau' - \tr)^2\over \dtr^2} \right\}. \eqnum\gapp $$
For standard recombination, $\tau_R \simeq .02\tau_0$ while the width of the
last scattering surface, $\dtr \simeq .1\tr$. For no recombination,
$\tr\sim .08\tau_0$ while $\dtr\sim.06\tau_0$.
\putfig\VISIBLE{visible.eps}{The visibility functions for standard
recombination [solid line] and no recombination [dashed line].}

Our goal is to find out how $\Dt$ depends on the ionization parameters,
$\tr$ and $\dtr$. With the approximation in Eq. \gapp\ we can go further and
solve explicitly for $\Dt$ in Eq. \simpsol. We now have
$$ \Dt(k,\mut,\tau_0) = 4\mu {1\over  \sqrt{\pi}\dtr}
\int_0^{\tau_0} d\tau' \vt(k,\tau')
e^{ik\mut(\tau'-\tau_0)}
	\exp\left\{
		- {(\tau' - \tr)^2\over \dtr^2} \right\}.\eqnum\apps $$
In a matter dominated Universe, $\vt$ grows as
$\tau$, so $\vt(\tau) = \vt(\tc) \tau/\tc$. We can also set the upper limit of
the
integral to infinity since the exponential is negligible for $\tau>\tau_0$.
Then,
$$ \Dt(k,\mut,\tau_0) = {4\mu \vt(k,\tc)\over  \sqrt{\pi}\tc \dtr}
e^{-ik\mut\tau_0} {1\over i}{\partial\over\partial(k\mut)}\int_0^\infty d\tau'
e^{ik\mut\tau'}
	\exp\left\{
		- {(\tau' - \tr)^2\over \dtr^2} \right\}.\eqnum\appso $$
where we have written $\tau'$ as $(1/i)\partial/\partial(k\mut)$.
The integral in Eq. \appso\ contains a lot of the physics we are after.
\putfig\integrand{integrand.eps}{Real part of the integrand of Eq.\ \appso.}
Figure \integrand\ shows the integrand for two limiting cases: (i) $k\mut\dtr
>> 1$ and (ii)
$k\mut\dtr << 1$. In the first case [large $k$], the scale of the perturbation
is much smaller than $\dtr$, the width of the surface of last scattering.  A
photon travelling through the last scattering surface travels through many
regions where $v$ is positive but an almost equal number of regions where $v$
is negative. Thus the total contribution to $\Dt$ is small on scales smaller
than the thickness of the last scattering surface. The second case comes about
either because $k$ is small [i.e. the scale of the perturbation is large] or
$\mut\sim 0 ~[\h p \cdot \h v \propto \vec p\cdot\nabla \delta \sim 0$, the
photon is travelling
perpendicular to the direction in which the perturbation is changing].
Perturbations on large scales {\it do} make a large contribution to $\Dt$ since
there is no cancellation through the last scattering surface. Similarly, there
is no cancellation [in the integral] if the photon is travelling perpendicular
to the gradient of the perturbation. [Note though that Eq. \appso\ has a factor
of $\mu$
in front which does lead to a cancellation for such photons.]
In any event, the integral under discussion can be written down in terms of
Error
functions, but for our purposes we will make another approximation which will
simplify things further. The contribution of the integrand at the lower limit
$\tau' = 0$ is suppressed by a factor of $e^{-(\tr/\dtr)^2}$; for most
realistic ionization histories this will be a pretty small number, so we can
extend the lower limit all the way to $-\infty$ with little loss of accuracy.
The remaining integral is then the Fourier transform of a Gaussian, which is
itself a Gaussian, or
$$ \dd(k,\mu,\tau_0) = 4\mu \vt(k,\tc)\ {\tr\over\tc}\left[
	1 \ + \ i{k\mu\dtr^2\over 2\tr} \right]
e^{ik\mu(\tr-\tau_0)} e^{-(k\mu\dtr/2)^2} ,\eqnum\appdel $$
where the superscript indicates that this is the perturbation to the photons
caused by the Doppler effect.
Eq.\ \appdel\ clearly illustrates the damping on small scales as we have
discussed; the damping scale is of order $k\sim 1/\dtr$. Kaiser (1984) pointed
out that since the velocity is parallel to ${\bf k}$, there is an extra factor
of $\mu$ in front of Eq. \appdel. Thus, for example,
$\delta\rho_\gamma/\rho_\gamma = (1/2)\int_{-1}^1 d\mu\
\dd(k,\mu,\tau_0)\propto 1/(k\dtr)^2$. This suppression of
$\delta\rho_\gamma/\rho_\gamma$ would be averted if the source term had a
component perpendicular to ${\bf k}$. The other feature of Eq.\ \appdel\ which
is of interest is the factor of $\tr/\tc$ in front. This tells us that the
later the photons are in contact with the electrons, the greater is the Doppler
kick they get. But this is what we expect: velocities grow with $\tau$, so
photons scattering off electrons at later times are seeing larger velocities.
This simple fact will have surprising consequences, namely, on large scales
[before damping sets in] a reionized Universe produces a {\it larger} signal in
the CMB than one which never reionizes!

What is the contribution of the Doppler effect to the present day anisotropy as
measured by the $C_l$'s? Referring back to Figure 2, we see that for $l < 30$
or so, the dominant
contribution to $C_l$ comes from the Sachs-Wolfe effect, for which $l(l+1)C_l$
is
constant. At larger $l$'s, the Doppler effect becomes relevant.
The reason for deriving an analytic expression for $\dd$ in terms of $\dtr$ and
$\tr$ is to see when the Doppler effect becomes important as we vary the
ionization history of the Universe. To calculate the $C_l$'s, we expand
$\dd(k,\mu,\tc)$
in a series of Legendre polynomials, defining
\be
\dd_l(k,\tc) = {1\over 2} \int_{-1}^1 d\mu \ P_l(\mu) \dd(k,\mu,\tau_0).
\ee
Then, $C_l$ is given by:
\bea\cldefin{
\cldop &= {V\over 8\pi} \int_0^\infty dk\ k^2 \ \langle\left\vert \dd_l(k,\tc)
\right\vert^2\rangle\cr\vs
&= {V\over 8\pi} \int_0^\infty dk~k^2~
\langle\Bigg\vert
{1\over2}\int_{-1}^1 d\mu P_l(\mu)
4\mu \vt(k,\tc)\left[
	{\tr\over\tc} \ + \ i{k\mu\dtr^2\over 2\tc} \right] \cr\vs
&\qquad\qquad\qquad \times \ e^{ik\mu(\tr-\tau_0)} e^{-(k\mu\dtr/2)^2}
\Bigg\vert^2\rangle.
}
Here we are using the normalization of Efstathiou, Bond, $\&$ White (1992). [A
good way to check normalization is to insure that $C_2 =
(8/\pi\tc^4)\int_0^\infty dk j_2^2(k) P(k)/k^2$, where $P(k)$ the power
spectrum is normalized to COBE, so on large scales is proportional to $C_2 k$.]
As a first approximation to $\cldop$, we'll assume that only $k < 2/\dtr$
contribute to the $k-$ integral due to damping, and that for these values of
$k$ we can set the exponential $e^{-(k\mu\dtr/2)^2}$ to one. Then the integral
over $\mu$ is simply the derivative of a spherical Bessel function:
$$ {1\over2}\int_{-1}^1 d\mu \mu P_l(\mu) e^{ik\mu(\tr-\tau_0)}
= {1\over i}{\partial\over\partial (k(\tr-\tau_0))} {j_l(k(\tr-\tau_0))
\over (-i)^l}, \eqnum\ptobess $$
and similarly the term with $\mu^2$ becomes the second derivative of a
spherical Bessel function. Also the continuity equation tells us that ${\bf
\vt}(k,\tc) = -i\bk \dot{\delt}(k,\tc)/k^2 = -2i\bk\delt(k,\tc)/k^2\tc$, where
$\delt$ is the fractional change in density of the electrons, assumed equal to
that of the rest of the matter. Thus, the ensemble average is:
$$ <\vert \vt(k,\tc) \vert^2> = {4\over (k\tc)^2}~<|\delt(k,\tc)|^2>
= {4P(k)\over V(k\tc)^2} .\eqnum\contin $$
Here $P(k)$ is the COBE normalized power spectrum;for CDM\rc\EBW,
$
P_{CDM}(k) = {3\pi\over 2} C_2 \tc^4 kT^2(k)
$
where $C_2= 4\pi(Q_{RMS}/T_0)^2/5 = 9.7\times 10^{-11}$ and $T$ is the transfer
function.

We now have our final result for $\cldop$ in terms of a simple one dimensional
integral:
\belabel\cldfin
\cldop \simeq {8 \over \pi} \left( {\tr\over\tc}\right)^2
\
\int_0^{2\tc/\dtr} dz~ {P(k=z/\tc)\over\tc^3}\ \left[
			{d j_l(z)\over dz}
- {\dtr^2\over 2\tc\tr} z {d^2j_l(z)\over dz^2} \right]^2 .\eqnum\clfdop $$
\putfig\clappvsex{clappvsex.eps}{$l(l+1)C_l$ for standard recombination and no
recombination. The results of a full numerical solution of the coupled
Boltzmann equations
[solid lines] are compared with the approximate solution given in Eq.\ \clfdop\
[dashed lines].}
Figure \clappvsex\ shows this approximate result for $\cldop$ added to the
Sachs-Wolfe result of $\clsw = 6C_2/l(l+1)$ for standard recombination and no
recombination. Also shown are the results of numerically integrating the full
set of Boltzmann equations [including perturbation to the metric, CDM and three
species of massless neutrinos] for these two ionization histories. Apparently
the approximate result of Eq.\ \cldfin\ gives a very good qualitative picture
of the results we are after: (i) the Doppler peak {\it is} larger on large
scales [small $l$] and (ii) the damping {\it does} take place at larger scales
than in the case of standard recombination. Clearly, though, for an accurate
quantitative analysis the equations must be solved numerically.
One final quantitative point: Having emphasized the fact that $C_l$ is larger
in a no-recombination Universe at low $l$, we should point out that it's not
{\it that} much larger. The difference never exceeds $30\%$ for any $l$. Thus
we'll see in Section 7 that the expected $\Delta T$ [which goes as the square
root of $C_l$] for no-recombination is roughly the same on these scales as for
standard recombination. All our work in this section has been on the linear
terms in the Boltzmann equation. We now turn to second order terms.

\newsec{Second Order Equation}

In this section we write down the second order equation which follows from our
general Boltzmann treatment in Section 2. It turns out that there are numerous
terms. We can immediately throw out one class of terms in Eq. \allcoll\ though.
The terms coupling $f^{(1)}$ and $v$, i.e. those in Eq. \edelv, are nominally
second order. In practice, though, we have seen that $f^{(1)}$ [or its
equivalent $\Delta^{(1)}$] is very small on small scales due to damping. So we
we can neglect $c^{(2)}_{\Delta v}$ in Eq. \allcoll. The remaining terms can be
manipulated as were the first order terms in Section 3. The result for the
second order collision term is
\def\nn{\cr\vs}
\bea\ctwojay{
C^{(2)}(\bp) &= n_e\sigma_T
\Biggl\{-p {\partial f^{(0)} \over \partial p}\delta_e \mu v
\ + \ f^{(2)}_0
+ {1 \over 2}f^{(2)}_2 P_2(\mu)
-f^{(2)}\nn\qquad
& +v^2p {\partial f^{(0)} \over \partial p}
(\mu^2 + 1) +
v^2p^2{\partial^2 f^{(0)} \over \partial p^2}
({11 \over 20}\mu^2 + {3 \over 20}) \nn \qquad\qquad &
+ { 1 \over m_e^2}{\partial \over \partial p}
\left[ p^4 \left(T_e {\partial f^{(0)} \over \partial p} + f^{(0)}
(1 + f^{(0)})
\right)
\right] \Biggr\}
}
The first term on the right hand side arises from writing the electron
density as $n_e=\bar n_e(1+\delta_e)$ and then multiplying the first
order collision terms (Eq. \efirst) by $\delta_e$ [note that we have dropped
all terms with $f^{(1)}$'s in them].  This is the so-called
``Vishniac term''.
The remaining terms arise from the second order expansion in the
collision integral, Eq. \allcoll.
At second order, electron-photon scattering is not purely elastic.
Energy is transferred between the electrons and photons,
inducing spectral distortions.   In the limit
of no anisotropy
only the terms on the last line, the Kompaneets terms which
give rise to spectral Sunyaev-Zel'dovich distortions, survive.
Since the Kompaneets terms do not induce anisotropies, we shall neglect them.
Finally, since we are interested in scales much smaller than the
horizon, non-Newtonian gravitational effects can be ignored.

It would be nice to define a $\Delta^{(2)}$ just as we did $\Delta^{(1)}$ in
Eq. \defdel\ so that the second order equation is momentum independent. This is
impossible though, because the second order equation is {\it momentum
dependent}! Specifically the $v^2$ terms in Eq. \ctwojay\ are momentum
dependent. Thus in general $\Delta^{(2)}$ will depend on momentum. There are
two ways to deal with this: Either we can go ahead and try to solve the full
equation for $\Delta^{(2)}(k,p,\mu)$, or we can integrate out the momentum
dependence. Let us first take the easy way out, and later we will see if it is
necessary to include the full momentum dependence. To do the integration we can
define
\be
\Delta^{(2)} \equiv {\int dp p^3 f^{(2)} \over \int dp p^3 f^{(0)} },
\ee
as in Eq. \altdef. Then
integrating Eq. \ctwojay\ leads to the second
order equation:
\belabel\ebh
{\partial \Delta^{(2)} \over \partial \tau }  + a\hat p^i  {\partial
\Delta^{(2)} \over \partial x^i } =
a \bar n_e \sigma_T
\Biggl\{
4 \delta_e \h p \cdot\bv +
v^2 \bigg( 7 + 15 \big(\h p\cdot\h v\big)^2 \bigg) - \Delta^{(2)}
 \Biggr\},
\ee
where we have dropped the $\Delta^{(2)}_0$ and $\Delta^{(2)}_2$ terms on the
right hand side, as these should be irrelevant on small scales.
On physical grounds, we can argue that the only important second order term
is the ``Vishniac'' term.  The Vishniac term is proportional to $\delta_e v$
whereas the other terms are proportional to $v^2$.  By the continuity equation,
though, $v \sim \delta_e/k\tau$. Since $k\tau$ is quite large ($>100$) for the
scales of interest, the Vishniac term should dominate.

Are the other terms in Eq. \ebh\ completely irrelevant? Not necessarily. Recall
that we integrated out the frequency dependence of the velocity squared terms
in Eq. \ctwojay. Even if the final value of $<(\Delta T/T)^2>$ induced by these
terms was a factor of $100$ smaller than the leading terms, we might be able to
pick them out because of the frequency dependence. [A recent example of the
process of separating out a signal from a much
more powerful source with a different ``spectral index,'' i.e. frequency
dependence, can be found in Meinhold {\it et al.} (1993).] So we choose to keep
these additional terms a little longer.

\newsec{Second Order Contribution to ${\Delta T \over T}$ }
In this section we will follow the treatment of Efstathiou (1988) in solving
the second order equation. We can rewrite Eq. \ebh\
as
\belabel\ebhone
\dot{\tilde{\Delta}}^{(2)} + ik\mu \Dtwo =
\bar n_e \sigma_T a
\left[ \sdv(\bk,\tau) + \svv(\bk,\tau)- \Dtwo \right]
\ee
where the Vishniac source term is
\bea\eqsdv{
\sdv(\bk,\tau) &\equiv 4\hp \cdot \sum_{\bkp}
\bvt(\bkp,\tau)\tilde\delta_e(\bk-\bkp,\tau) \nn
&= \left( {\tau\over \tau_0} \right)^3 {8i\over \tau_0} \sum_{\bkp}  {\hp \cdot
\bkp\over k'^2}\
\delt(\bkp,\tau_0)\delt(\bk-\bkp,\tau_0)
}
and the source term quadratic in velocities is
\bea\eqsvv{
\svv(\bk,\tau) &\equiv \sum_{\bkp}
\Big[ 7\bvt(\bkp,\tau) \cdot \bvt(\bk-\bkp,\tau)\ +\ 15 \hp
\cdot\bvt(\bkp,\tau)
				\hp\cdot\bvt(\bk-\bkp,\tau) \Big]
 \nn
&= {-4 (\tau/\tc)^2\over \tau_0^2}  \  \sum_{\bkp}
{ 7 \bkp\cdot(\bk-\bkp) \ + \ 15\hp \cdot \bkp \hp \cdot (\bk - \bkp)\over
|\bk-\bkp|^2 k'^2}\
\delt(\bkp,\tau_0)\delt(\bk-\bkp,\tau_0).\cr &
}
Here we have Fourier transformed from position space to
momentum space. We have also used the facts that (i) the velocities are first
order, and are related
to the baryon density perturbations through the equation of continuity,
and (ii) the time dependence of the perturbations is given by $\delt(\tau) =
\delt(\tc)({\tau \over \tc})^2$ in a matter dominated Universe.

The second order equation is of the same form as the first order equation,
whose solution we wrote down in Eq. \simpsol. By analogy, the solution to the
second order equation is
\be
\Dtwo(k,\mu,\tau_0) = e^{-ik\mu\tau_0} \left[ \sdv(\bk,\tau_0) I_1(k\mu\tc)
+ \svv(\bk,\tc) I_2(k\mu\tc) \right]
\ee
where the time
integrals are
\bea\deftimeint{
I_1(k\mu\tc) &=\int_{0}^{\tc} d\tau' \Bigl({\tau' \over \tc}\Bigr)^3
g(\tc,\tau')e^{ik\mu\tau'} \nn
I_2(k\mu\tc) &=\int_{0}^{\tc} d\tau' \Bigl({\tau' \over \tc}\Bigr)^2
g(\tc,\tau')e^{ik\mu\tau'}
}
with $g$ our old friend, the visibility function.
We encountered an integral of this form [only the powers of $\tau'$ differ] in
section 4 when analyzing the first order equation. The main lesson we learned
was that integrals of this type are strongly damped unless $k\mu$ is small. On
small scales [large $k$] this means that the integrals are non-negligible only
if $\mu\simeq 0$. Thus the main contribution to $\Dtwo$ can be found by
evaluating the $\sdv(\tc)$ and $\svv(\tc)$ at $\mu=0$. Before doing this, let's
go back to Eq. \simpsol. There the source term was proportional to $\mu$, so
the contribution of the linear source term was greatly suppressed. Vishniac's
profound observation and the origin of the dominance of second order terms is
that the second order terms $\sdv$ and $\svv$ do {\it not} vanish at $\mu=0$.

For the very small angle experiments where second order effects are important,
there is a simple formula for the $C_l$'s:
\belabel\clder
C_l = {lV\over 32\pi \tau_0^3} \int_{-1}^1 d\mu\ \DDL
\ee
where $V$ is the volume
which will drop out at the end of the calculation.
To derive this, start with the small angle formula \rc\DOROSH:
\be
C(\theta,\sigma)=
 V\int_{0}^{\infty}{dk\ k^2\over64\pi^2} \int_{-1}^{+1}d\mu(1-\mu^2)
\DD J_0\left(kR_c\theta(1-\mu^2)\right)\ \exp\left[-(kR_c\sigma)^2\right]
\ee
where $\theta$ is the angle between the two observing direction; $\sigma$ is
width of the Gaussian beam observing the temperature
differences; and $R_c$ is the comoving distance to the last scattering surface,
$R_c \approx \tau_0$ in a flat universe with $z_c \gg 1$. Since $\DD$ is highly
peaked around $\mu=0$, we can set $\mu=0$ everywhere else
in the integrand. Meanwhile the product of $J_0$ and the exponential
suppression is just the experimental window function referred to in Eq. (1.1).
Changing variables to $l=k/\tc$ leads to
\be
C(\theta,\sigma)={V \over 64\pi^2 \tc^3} \int_{0}^{\infty}dl\ l^2
\int_{-1}^{+1} d\mu\
\DDL
W_{l, expt}.
\ee
Comparing this to the continuous form of Eq.\ (1.1) leads to Eq. \clder.
Let us now calculate the ensemble average $\DDL$:
\bea\ddexp{
\DD &= <|\sdv|^2> |I_1|^2
+ <\sdv\svv^*>I_1I_2^* + <\sdv^*\svv>I_1^*I_2\nn&\qquad\qquad\qquad
+ <|\svv|^2> |I_2|^2.
}
We argued above that $\sdv/\svv < 10^{-2}$, so the last term here is suppressed
by a factor of at least $10^{-4}$ relative to the first and we can safely drop
it. The only terms with $\svv$ that might be interesting are the cross terms,
so we will keep these. Upon squaring $S$, we encounter double sums, say over
$\bkp$ and $\bkpp$; thus we need the identity
\be
< \delt^*(\bkp) \delt^*(\bk-\bkp) \delt(\bkpp) \delt(\bk-\bkpp) >
	=\ {P(\bkp) P(\bk-\bkp)\over V} \left[ \delta_{\bkp,\bkpp} +
			\delta_{\bkp+\bkpp,\bk}\right]
\ee
where $P$ is the power spectrum today and $\delta$, the Kronecker delta.
We can use this expression to expand Eq.\ \ddexp\ as
\beanolabel{
\DD &= {64\over V^2\tc^2}
\sum_{\bkp,\bkpp} \ P(\bkp) P(\bk-\bkp) \left[ \delta_{\bkp,\bkpp} +
			\delta_{\bkp+\bkpp,\bk}\right]
\ {\h p \cdot\bkpp\over k''^2}\nn &\times
\Bigg\{
	{\h p\cdot\bkp\over k'^2} |I_1|^2
+
	{ 7 \bkp\cdot(\bk-\bkp) \ + \ 15\hp \cdot \bkp \hp \cdot (\bk - \bkp)\over
k'^2|\bk-\bkp|^2\tc} {\rm Im}(I_1I_2^*) \Bigg\}
.}
To go further we use the Kronecker deltas to get rid of the $\bkpp$ sum and
change the $\bkp$ sum into an integral via $\sum_{\bkp} \rightarrow
V\int d^3k'/(2\pi)^3$. Furthermore, due to the damping effect we can set
$\h p\cdot \bk=\mu=0$ everywhere except in the time integrals ($I_1,I_2$).
Thus,
\beanolabel{
\DD &= {64\over V\tc^2 (2\pi)^3}
\int {d^3k'\over k'^2} \ P(\bkp) P(\bk-\bkp)\ \h p \cdot\bkp\
\left( {1\over k'^2} - {1\over \kmkp^2} \right)
\nn &\times
\Bigg\{
	\h p\cdot\bkp |I_1|^2
 +
	{ 7 \bkp\cdot(\bk-\bkp) \  - \ 15(\hp \cdot \bkp)^2 \over |\bk-\bkp|^2\tc}
{\rm Im}(I_1I_2^*) \Bigg\}
.}
With no loss of  generality we orient  our coordinate system such that:
\be
\bk \cdot \hp = 0; \quad
\bk \cdot \bkp = k k' x; \quad
\bkp \cdot \hp = k'\sqrt{(1-x^2)} {\rm sin} \phi
\ee
Then
\bea\ddexptwo{
\DD &= {64\over V\tc^2 (2\pi)^3}
\int_0^\infty dk'\ \int_{-1}^1 dx\ \int_0^{2\pi}d\phi\ \
P(k') P\Big( (k^2+k'^2-2kk'x)^{1/2}\Big)\nn &\quad\times
k'\sqrt{1-x^2} \sin\phi\
\left( {1\over k'^2} - {1\over \kmkpsq} \right) \nn
&\qquad\times\Bigg\{
	k'\sqrt{1-x^2} \sin\phi |I_1|^2
\ +\
	{ 7 (kk'x-k'^2) \ - \ 15k'^2(1-x^2)\sin^2\phi\over (\kmkpsq)\tc} {\rm
Im}(I_1I_2^*) \Bigg\}.\cr&}
We now see that the cross terms -- those with $I_1I_2$ -- vanish once the
$\phi$ integral is done (Hu {\it et al.} 1993). Thus we conclude that {\it the
only second order term of any
significance is the Vishniac term}.
Plugging this expression into Eq.\ \clder\ yields
\be
C_l={l  J P^2(k=l/\tc) \ K(k=l/\tc)\over 2\pi^2 \tc^6}
\ee
where
\be
K(k)={1 \over k}
\int_{0}^{\infty} dk' \int_{-1}^{+1} dx
{P(k')P\Big((k^2+k'^2-2kk'x)^{1/2}\Big) \over P^2(k)}
{(1-x^2) (k^2-2xkk') \over (k^2+k'^2 -2xkk')}
\ee
and
\be
J\equiv {k\tc\over2\pi}\int_{-1}^1 d\mu\ \vert I_1(k\mu\tc) \vert^2.
\ee
For CDM, $K$, the integral over the power spectrum, can be approximated by
$K(l) \simeq -0.1+ 1.1(l/1000)$ for $1000<l<5000$, with a maximum error of
$5\%$. [The integral $K$ is half of the integral Efstathiou (1988) calls $I_2$.
It's also much easier to compute numerically. To prove the identity, one uses
the invariance of the integrand under $\bkp \leftrightarrow \bk-\bkp$.]

\putfig\VISH{vish.eps}{$C_l$'s induced by the second order Vishniac term for
CDM with no recombination.}

The integral $J$ contains all the information about the ionization history, but
it also appears to depend on wavenumber $k$. Efstathiou (1988) made the
intriguing observation that for most reasonable ionization histories, $J$ is in
fact independent of $k$. To see this, we note that the integrand is sharply
peaked around $\mu=0$ so there should be very little error introduced if we
extend the limits of integration all the way to $\pm\infty$. Then,
\bea\jint{
J & \simeq {1\over 2\pi} \int_{-\infty}^\infty d(k\mu\tc)
\int_0^{\tc}
	d\tau'\  \Bigl({\tau' \over \tc}\Bigr)^3
g(\tc,\tau')e^{ik\mu\tau'}
\int_0^{\tc}
	d\tau''\  \Bigl({\tau'' \over \tc}\Bigr)^3
g(\tc,\tau'')e^{-ik\mu\tau''} \nn
&=\int_0^{\tc}d\tau' \Bigl({\tau' \over \tc}\Bigr)^6
g^2(\tc,\tau') \tc}
where the last equality follows immediately since the $(k\mu\tc)$ integral
yields a delta function in $\tau'-\tau''$. All the information about the
ionization history is in the
integral $J$. From Eq.\ \jint, we see that the integrand is heavily towards
late times by the $\tau'^6$ factor. Thus ionization histories wherein
the visibility function is peaked at late times --  that is, scenarios in which
the Universe is re-ionized -- are most likely to produce appreciable secondary
anisotropies. For standard recombination,  $J = 1.7\times 10^{-8}$ while $J=
7.3\times 10^{-5}$ for no recombination.

Figure \VISH\ shows the $C_l$'s for CDM with no recombination. The shape of the
curve is exactly the same for other ionization histories, only the amplitude,
which is determined by $J$, drops. Nonetheless, since the integrand of $J$ is
heavily weighted towards late times, even relatively late re-ionization
produces a comparable signal. The other feature of note in Fig. \VISH\ is the
amplitude, which is small. We'll see in the next section how this translates
into an expected $\Delta T$ in a given experiment, but clearly it will be
difficult to detect these secondary anisotropies in CDM. Other cosmologies, in
particular baryon isocurvature models, are more likely to produce a large
secondary signal due to the Vishniac effect\rc{\BEFS; \Efstathiou; Hu {\it et
al.} 1993}. Although we won't calculate the signal in these models here, we
stress that our conclusion that Vishniac's term is the only important second
order one applies in general.

\medskip\goodbreak\global\advance\countsec by 1
{\parindent0pt{\bf \number\countsec. Results and Discussion}}
\nobreak\medskip\nobreak\counteqn = 0

Now that we have all the physics under our belts, it is time to probe different
ionization histories. In this section we present the $C_l$'s for a variety of
ionization histories and convolve them with the filter functions shown in
Figure \FILTERS\ to obtain the predicted signal in these experiments.

\putfig\VISALL{visall.eps}{The visibility functions for the ionization
histories we will discuss in this section. Shown for comparison are the
standard recombination history and the fully ionized history. The small numbers
alongside each line indicate the redshift of complete ionization.}

Figure 7 shows the visibility functions for the histories we will discuss. Note
that the ionization in these particular histories is gradual so that even the
history wherein ionization takes place at $z=10$ differs from the standard
recombination. The advantage of these particular histories is that they were
generated in a consistent manner; i.e. a source of photons was injected
continuously into the medium and the effects of recombination and electron
heating were taken into account\rc\DDM. The ionization history of the Universe
{\it could} have been as sketched in any of the lines in Figure 7.
The disadvantage of these histories is that in all of them re-ionization is
gradual. We would like to be able to say something general about re-ionization
without reference to these specific histories. Therefore, referring to them by
the epoch of complete re-ionization is {\it not} a good idea.

It is more useful to discuss the {\it cumulative visibility function}:
$\int_\tau^{\tau_0} d\tau' g(\tc,\tau')$;
this is shown in Figure 8. The cumulative visibility function is the
probability that a photon has last scattered after a given time. Thus, for a
no-recombination Universe, Figure 8 shows that almost all photons scattered
after $\tau = .05\tc$. By contrast, the cumulative visibility function is only
equal to $.005$ at $\tau = .05\tc$ for the standard recombination history. The
other ionization histories lie somewhere in between.  A nice feature of this
number [the cumulative visibility function at $\tau = .05\tc$] is that it
characterizes each ionization history in an easily understandable way. Further,
it lies between $0$ [standard recombination] and $1$ [no recombination] for all
reasonable ionization
histories.

\putfig\CUMVIS{cumvis.eps}{The cumulative visibility functions for the
ionization histories we will discuss in this section. The numbers by each
denote the value of the cumulative visibility function when $\tau = 0.05\tc$.}

\putfig\CLALL{clall.eps}{The $C_l$'s for different ionization histories ranging
from standard recombination to no recombination. Again the numbers give the
value of the cumulative visibility function at $\tau=.05\tc$.}

For each of these histories we have calculated the expected signal in the CMB
in the form of the $C_l$'s.
Figure 9 shows the results of the numerical integration of the Boltzmann
equations. Working our way down from standard recombination, we see that if the
cumulative visibility function is small at $\tau=.05\tc$ [e.g. the curves
labeled $.10$,$.28$, and $.59$], the effect of reionization is to damp out the
primary anisotropies generated early on. The more effective the reionization
[as parametrized by a larger cumulative visibility function], the more
significant is the damping of the peak at $l=200$. Indeed, if most photons
scattered late [$\tau > .05\tc$], then this peak goes away completely [e.g. the
curves labeled $.88$ and $.98$], as we expect from our discussion in section 4
of the no recombination case. At the same time, though, secondary anisotropies
-- those generated at late times when velocities were larger -- become
important as reionization becomes more efficient. So for $l<50$, $C_l$ is
actually {\it larger} in a Universe which reionizes early.

We now convolve these $C_l$'s with the filter functions for the experiments
plotted in Figure \FILTERS. We are interested in the expected signal, as
defined in Eq. \cldef. So we plot $\langle \Delta T_{\rm expt}^2 \rangle^{1/2}$
as a function of the reionization parameter.

{\it The expected signal in the Tenerife experiment is virtually independent of
the ionization history.} We see a small increase in the signal for
no-recombination vs. standard recombination since the Tenerife filter samples
part of the Doppler no-recombination peak. But the difference is very small, of
order $5\%$, certainly too small to be meaningful at present.

\putfig\SIGNAL{signal.eps}{The expected $<\Delta T^2>^{1/2}$ in a variety of
experiments as a function of the reionization parameter, the cumulative
visibility function at $\tau=.05\tc$. Small values of the cumulative
visibility function correspond to standard
recombination; values close to one correspond to no recombination.
The signal in OVRO includes the second
order Vishniac effect.}

The South Pole 91 filter is situated between the standard recombination peak at
$l\sim 200$ and the no-recombination peak at $l\sim 50$. Since the former peak
has a larger amplitude,
the signal in SP91 is larger if the Universe had a standard recombination
history than if it never recombined. It is interesting to note though that the
curve is not monotonic: If the cumulative visibility function at $\tau =
.05\tc$ lay between $0.5$ and $1$, the Doppler peak at $l\sim 200$ is
significantly depleted and the secondary peak at $l\sim 50$ has not built up to
its maximum value. Thus the expected signal in medium scale anisotropy
experiments is lowest if roughly half the photons last scattered
at the standard $z\simeq 1100$ and the other half at $z\simeq 100$. The signal
can drop by as much as a factor of two from what would be expected in standard
recombination. Amusingly, the observed signal in all four channels {\it was}
roughly a factor of two below what would be expected, but this must be taken
with a grain of salt since there was probably a great deal of contamination
from foreground sources\rc{\GAIER; \US}.

The MAX experiment has a filter centered closer to the primary peak at $l\simeq
200$. Therefore, the signal drops even more precipitously when the Universe is
re-ionized and the primary peak is washed out. The signal can be a factor of
three smaller than in standard recombination. Note again the effect of the
secondary peak: as the cumulative visibility function reaches one, the
secondary peak and, therefore, the signal increases. {\it MAX and SP91 are both
very good probes of the ionization history, with the expected signal varying by
a factor of $2-3$ depending on the ionization history.}

{\it By far the best probe of ionization history is the Owens Valley Radio
Observatory receiver.} Note from Fig. \FILTERS\ that the OVRO filter does not
pick up the secondary peak at $l\sim 50$. So the signal due to first order
effects [pre-Vishniac] is completely negligible if the Universe never
recombined. In fact, the expected signal drops by a factor of order ten if the
Universe never recombined. This drop is large but it would be even more
dramatic if not for the Vishniac effect, which produces a small, but
non-negligible signal in the case of no-recombination when the primary signal
has been completely washed out.

To sum up, Fig. \SIGNAL\ gives meat to the common wisdom that the smallest
scale experiments are most sensitive to the ionization history of the Universe.

We started this investigation wondering whether a signal due to re-ionization
could be misinterpreted as a primordial signal. We now know how a signal due to
re-ionization differs from one due to standard CDM with standard recombination
history, and we are confident that current and future experiments will probe
such differences. What we have not done in this paper is explore how a signal
due to re-ionization differs from a signal in other variants of CDM. Most
troublesome are two variants most likely to be confused with re-ionization: (i)
models where the primordial spectrum is {\it not} Harrison-Zel'dovich $(n< 1)$
and (ii) models in which there are primordial tensor perturbations due to
gravity waves\rc\CRITTENDEN. Both of these have the feature that the signal on
large scales is the same as in standard CDM, while the signal on small scales
is smaller than the standard one. They share this feature with re-ionized
models. Therefore, distinguishing tilted or gravity-wave models from re-ionized
models is a challenging task for cosmologists.

It is a pleasure to thank Douglas Scott for conversations about this work and
for showing us the preprint of Hu {\it et al.} (1993) prior to its release. We
also thank Ed Bertschinger, Katherine Freese, Arthur Kosowsky, Albert Stebbins,
and Michael Turner for helpful discussions.
The work of SD was supported in part by the DOE and NASA
grant
NAGW-2381 at Fermilab.
\immediate\closeout\refout
	\medskip
	\centerline{\bf REFERENCES}
	\medskip
	\input reference.txa
\bye